\documentclass[prd,twocolumn]{revtex4}
\usepackage{graphicx}                                                                                                                                                                                                                             
\usepackage{amsmath}
\usepackage{amsfonts}
\usepackage{amssymb}
\usepackage{mathrsfs}
\usepackage{textcomp}
\usepackage{graphics}
\usepackage{caption}
\usepackage[titletoc]{appendix}
 \usepackage{subfig}
 \usepackage[utf8]{inputenc}

\begin{document}

\begin{abstract}
In this paper we consider a model of quintessential inflation based upon inverse hyperbolic potential. 
We employ curvaton mechanism for reheating 
which is more efficient than
gravitational particle production; the mechanism complies with nucleosynthesis constraint due to relic gravity waves. We obtain a lower bound on the coupling constant $g$ that governs the interaction of curvaton with matter fields. We 
study curvaton decay before domination and decay after domination and plot the allowed region in the parameter space in both cases.
\end{abstract}

\title{ Quintessential Inflation and curvaton reheating}
\author{Abhineet Agarwal} 
\affiliation{ Centre for theoretical physics Jamia Millia Islamia, New Delhi}
\author{Sabit Bekov}
\email{
sabit bekov <ss.bekov@gmail.com>} \affiliation{ Eurasian  International
Center for Theoretical Physics, Eurasian National University, Astana
010008, Kazakhstan}
\author{Kairat Myrzakulov}
\email{kmyrzakulov@gmail.com} \affiliation{ Eurasian  International
Center for Theoretical Physics, Eurasian National University, Astana
010008, Kazakhstan}                                                                             
\affiliation{Jamia Millia Islamia}
\date{\today}    

\maketitle

\section{Introduction}

The standard model of the Universe has many problems which need to be addressed both at early and at late times. 
A few examples of the most pertinent problems are the flatness problem, the monopole problem, the horizon problem and the age
crisis of the Universe.These
problems can be solved by introducing an early phase of accelerated expansion, called inflation, together with late
time cosmic acceleration. 
Recent data obtained from type Ia supernovae, CMB background and galaxy clustering suggest late time cosmic acceleration. 
 \cite{Abhineet,CCIsBack,GuthInflation}. 

Both inflation and late time acceleration are often studied separately but if we
build a theory in which both the above phenomena are implemented by using a single scalar field
then a unique cause could drive both. This is known as ''quintessential inflation'' and its advantage
is that it is efficient as it provides a common theoretical framework\cite{Wali1,Wali2,WoodOwen,QIPeebles}.  
In order to achieve this
unification the field must first evolve very slowly in order to drive inflation, 
then it should roll very quickly in order to exit from inflation. 
Then comes a time when the 
scalar field kind of disappears from the scene and its
evolution no longer plays a prominent role in the history of the Universe.
It then reemerges to play a key role in driving late time cosmic acceleration.
Thus inflation is reborn as late time cosmic acceleration.

In order to model the desired behaviour of the scalar field, we usually describe the dynamics of the scalar field with
the help of a potential $V( \phi )$ which is unknown so far and whose functional form is 
highly debatable \cite{SineHyperbolicInflation}. But the 
potential should have certain characteristics in order to achieve the desired result. It should be shallow at early
times in order for slow roll to commence.
For inflation to end the shallow behaviour should be followed by the potential falling off steeply. A scaling
solution is thus obtained where the energy density of field changes in exactly the same way as that of the background.            
The potential should then have certain late time characteristics giving rise to scaling behaviour.  Most generic
potentials do not change their shape frequently enough to be shallow early on, then steep and then again shallow at
late times. They are either shallow at early times and steep thereafter or vice versa. The first class
of models could give rise to a viable scenario of quintessential inflation if a suitable mechanism was constructed
in order to exit from the scaling regime in order to obtain late time acceleration. The second class
requires extra damping at early times in order to facilitate inflation but fails as it predicts a high
tensor to scalar ratio of perturbations and hence these models do not satisfy observational constraints.

We stick to the first case thus. In this case we exit from the scaling regime and obtain late time acceleration by 
coupling the field non minimally to matter, generally neutrinos as they become non relativistic at late times. This
creates a minimum in the field potential at late times without destroying the matter phase. 

Such a model with inverse hyperbolic type potential was recently studied and it was shown that coupling to massive
neutrino matter led to late time acceleration \cite{Abhineet} . The instant preheating mechanism with Yukawa interaction was considered
and bounds on the coupling constants were obtained using the nucleosynthesis constraint on relic gravity waves produced
during inflation. Bounds on the reheating temperature were also obtained. 

In this paper the curvaton mechanism is used to reheat the Universe and generate the large -scale curvature
perturbations in our Universe, which is the dominant cause of structure in the Universe. 

In the curvaton mechanism, the inflaton field only drives inflation. During the inflationary era the energy density
of the curvaton field is sub-dominant and is not diluted by the expansion. After horizon exit, the quantum fluctuations
of the curvaton field get converted into classical perturbations with a flat spectrum. This corresponds to
isocurvature perturbations. The moment inflation ends the energy density of the curvaton field becomes
(pre) dominant. Then the isocurvature perturbations are converted into adiabatic yielding curvature perturbations which
are fairly large in size. The curvaton then decays into conventional matter forcing the perturbations to stay adiabatic.
The curvaton is thus responsible for all the present material in the Universe and also for the Large Scale Structure of
the Universe 
\cite{CurvatonBrane,LythUngarelliWands,LythWands,FengLi,BartoloLiddle,MartinSloth,ByrnesCortLiddle,
MazumdarRocher,EnqvistLernerTaanila,GongKitajimaTerada,ChingangbamHuang}.

The advantages of curvaton reheating is that it is efficient unlike gravitational particle production and 
it solves many problems that are associated with various methods of reheating, 
like instant preheating, fermion preheating or those
based on inflaton decay 
\cite{InstantPreheating,InstantPreheating2,InflationafterPreheating,Campos1,Campos2,
FermionPreheating,Starobinsky,Starobinsky2}. One such problem in literature is the famous 
$\eta$ - problem \cite{DimopoulosEtaProblem,DimopoulosLyth}.
Introducing the curvaton makes it easier to construct sensible models of slow - roll inflation. With curvaton
reheating a higher reheating temperature can be obtained. 
Curvaton models predict a small value of tensor-to-scalar ratio \cite{KLTY}.
Also in this scenario gravitational waves of smaller
wavelength do not have a larger amplitude than desired and hence such waves do not dominate.
Curvaton models can also predict the duration of the inflationary 
period \cite{TorradoByrnesHardwickVenninWands}.

In this paper the same model with inverse hyperbolic type potential as discussed above is considered with curvaton as reheating mechanism.

\section{The Inflaton Field}

In this section we describe the inflaton field, a real scalar field denoted by $\phi$. 
The field does not couple much to other constituents of the Universe - matter, radiation and neutrinos.
Its dynamics is governed by the potential,

\begin{equation}
V (\phi) = \frac{V_0}{\cosh \Big( \frac{\phi^n}{\lambda^n} \Big)} 
= \frac{V_0}{\cosh \Big[ \beta^n \Big( \frac{\phi}{M_{pl}} \Big)^n \Big]} \label{potential}          
\end{equation}

,where $V_0$, $\lambda$ are free parameters. We consider a 
flat FRW background, with a metric,

\begin{equation}
ds^2 = - dt^2 + a(t)^2 \delta_{ij} dx^i dx^j.
\end{equation}
Here $\lambda = \alpha M_{pl}$ and $\beta = \frac{1}{\alpha}$, where $\alpha$ and $\beta$ are dimensionless parameters 
and $n$ is an integer. The following action is considered for the resulting dynamical system,
\begin{align}
\label{action}
S &= \int{d^4 x \sqrt{-g} \left[ \frac{M_{pl}^2}{2} R - \frac{1}{2} {\partial^{\mu} \phi} {\partial_{\mu} \phi} - V (\phi) \right]}
+S_m  \nonumber \\ & +S_{\nu}+ S_R.
\end{align}
 $S_m$, $S_\nu$, $S_R$ represent the actions for standard matter, massive neutrino matter and radiation respectively and all three
play a pivotal role in the post inflationary era.
The Friedmann equations for the  action (\ref{action}) in flat FRW geometry reduce to,
\begin{equation}
3 H^2 M_{pl}^2 = \rho_m + \rho_r + \frac{1}{2} {\dot{\phi}}^2 + V (\phi) \label{Frone},
\end{equation}
and
\begin{equation}
\left( 2 \dot{H} + 3 H^2 \right) M_{pl}^2 = - \frac{1}{3} \rho_r - \frac{1}{2} {\dot{\phi}}^2 + V (\phi) \label{Frtwo}  .
\end{equation}
The equation of motion for the scalar field is,
\begin{equation}
\ddot{\phi} + 3 H \dot{\phi} + \frac{d V}{d \phi} = 0.
\end{equation}
The slow roll parameters for a potential $V(\phi)$ are defined as usual \cite{QIPlanck2015SamiWali,LythontheHilltop},
\begin{equation}
\epsilon = \frac{M_{pl}^2}{2} {\left( \frac{1}{V} \frac{d V}{d \phi} \right)}^2 \label{epsilon},
\end{equation}
and
\begin{equation}
\eta = \frac{M_{pl}^2}{V} \frac{d^2 V}{d {\phi}^2} \label{eta}.
\end{equation}
The end of inflation is marked by,
\begin{equation}
\epsilon |_{\phi = \phi_{end}} = 1 \label{end} ,
\end{equation}
where ''end'' represents the value at the end of inflation.
Let us consider a period which begins when the modes cross the horizon and ends with the end of inflation. 
Then the number of e-foldings
during this period is given by \cite{QIPlanck2015SamiWali,LythontheHilltop},
\begin{align}
N(k) &= M_{pl}^{-1} \int_{\phi_{end}}^{\phi}{\frac{d \phi}{\sqrt{2 \epsilon(\phi)}}} \\
&= \frac{1}{M_{pl}^2} \int_{\phi_{end}}^{\phi}{\frac{V(\phi')}{V'(\phi')}d \phi'} \label{efoldings}  .
\end{align}
The tensor to scalar ratio $r$ is given by,
\begin{equation}
r = 16 \epsilon \label{ttsr},
\end{equation}
and the scalar spectral index $n_s$, which is defined as,
\begin{equation}
n_s - 1 = \frac{d(log P_R)}{d(log k)},
\end{equation}
where $P_R$ is the spectrum of curvature perturbations, is reduced to the form,
\begin{equation}
n_s = 2 \eta - 6 \epsilon + 1 \label{spectralindex}.
\end{equation}

Throughout the rest of the paper we use $n = 6$ and $\beta = 1$. For these values of $n$ and $\beta$, the
theoretical value of the spectral index $n_s$ is in agreement
with the Planck 2015 data up-to the $1 \sigma$  confidence level
and also the tensor-to-scalar ratio $r$ satisfies the Planck 2015 data, i.e., $r < 0.1$ \cite{planck2015}.

After inflation ends, the kinetic epoch begins and during this era the energy density of the inflaton field falls as,

\begin{equation}
\rho_{\phi} = \rho_{\phi}^{(kin)} \Big( \frac{a_{kin}}{a} \Big)^6 
\end{equation}

Thus the inflaton field describes what is known as ''stiff matter'' during the kinetic epoch.
We use ‘kin’ to label the value of the different quantities at the beginning
of the kinetic epoch.

Introducing dimensionless scalar field $\chi = \frac{\phi}{M_{pl}}$ in equation \ref{potential} one obtains, 

\begin{equation}
V = \frac{V_0}{\cosh \left( \beta^n \chi^n \right)} \label{dimensionlesspotential}        
\end{equation}

Now using the following values: 
$\chi_{in} = 0.44$, $\chi_{end} = 0.88$ and $V_0 = 4.64 * 10^{60} GeV^4$. 
Upon calculation we obtain initially, $V_i = V_{in} = 4.63988 \times 10^{60} GeV^4$
and also finally
$V_f = V_{end} = 4.18098 \times 10^{60} GeV^4$. Further using the relation, 
$H_{in}^2 = \frac{V_{in}}{3 M_{pl}^2}$ one obtains, 

\begin{equation}
H_{in}^2 = 2.68512 \times 10^{23} GeV^2
\end{equation}

and hence,

\begin{equation}
H_{in} = 5.18181 \times 10^{11} GeV 
\end{equation}

Also using, $H_{end}^2 = \frac{V_{end}}{2 M_{pl}^2}$, we calculate,

\begin{equation}
H_{end}^2 = 3.62932 \times 10^{23} GeV^2 
\end{equation}
 
and hence,

\begin{equation}
H_{end} = 6.02438 \times 10^{11} GeV 
\end{equation}

The amplitude of gravitational waves $h^2_{GW}$ is given by,

\begin{equation}
h^2_{GW} = \frac{H_{in}^2}{8 \pi M_{pl}^2} = 5.82139 \times 10^{-15} \label{amplitudegravitationalwaves}
\end{equation}

Gravitational waves behave as spin-less fields having no mass and
then their amplitude remains constant during inflation.
During the kinetic epoch, the energy density of gravitational waves evolves as,

\begin{equation}
\rho_g = \frac{64}{3 \pi} h_{GW}^2 \rho_{\phi} \Big( \frac{a}{a_{kin}} \Big)^2 
\end{equation}

During the moment when the energy density of the stiff scalar matter equals the energy density of the radiation
the energy density of the gravitational waves is given by, $\left( \rho_{\phi} = \rho_r \right)$ is,

\begin{equation}
\frac{\rho_g}{\rho_r} \Big|_{a = a_{eq}} 
= \frac{64}{3 \pi} h_{GW}^2 \Big[ \frac{a_{eq}}{a_{kin}} \Big]^2 \label{ratioofrhos}
\end{equation}

\section{The Curvaton Field}

The curvaton, denoted by $\sigma$,
is a real scalar field whose dynamics is governed by the potential -

\begin{equation}
U ( \sigma ) = \frac{m_{\sigma}^2 \sigma^2}{2}  \label{potentialcurvaton}
\end{equation}

and obeys the Klein Gordan equation,  

\begin{equation}
\ddot{\sigma} + 3 H \dot{\sigma} + m_{\sigma}^2 \sigma = 0
\end{equation}

Here $m_{\sigma}$ is the curvaton mass. During the inflationary period the curvaton field stays massless,
$m \ll H_f$ and it can be shown mathematically that under these conditions the curvaton field remains constant
during the inflationary era, $\sigma_f \simeq \sigma_i$ and also $\dot{\sigma}_f = 0$. Then we have,

\begin{equation}
U_f = \frac{1}{2} m_{\sigma}^2 \sigma_f^2 = \frac{1}{2} m_{\sigma}^2 \sigma_i^2 \label{Uf}        
\end{equation}

The subscripts $i$ and $f$ denote the values of the quantities at the beginning and end of inflation. 

When inflation ends, the kinetic regime begins, and it is during this era that the curvaton acquires a mass.
This happens at a time when $m \simeq H$.  We then get,

\begin{equation}
\frac{m}{H_{kin}} = \frac{a_{kin}^3}{a_m^3} \label{whencurvatonbecomesmassive}   
\end{equation}

Quantities labelled by the subscript $m$ indicate that they were evaluated at
exactly the same moment when the curvaton acquired a mass. Till this point, the curvaton
field stayed massless and then, $\sigma_m \simeq \sigma_i$.
We do not desire a period where the curvaton field drives inflation and for this to occur the 
the scalar
stiff-matter should dominate the Universe. 
Thus the energy of the curvaton field must be much less than that of stiff matter.
This leads to the equation,

\begin{equation}
\sigma_i^2 \ll \frac{3}{4 \pi} m_{pl}^2 \label{constraintonsigmasquare}
\end{equation}

where $\sigma_i$ is the initial value of the curvaton field.

\section{Constraining the Reheating Temperature}

\subsection{The Curvaton Reheating Mechanism}

If the Curvaton field, denoted by $\sigma$, decays into two fermions
$f$ and $\bar{f}$ during Curvaton Reheating then the Curvaton Reheating is governed by the reaction:

\begin{equation*}
\sigma \rightarrow f \bar{f} 
\end{equation*}

If the Lagrangian that governs this reaction is characterized by the coupling constant g, then the decay width 
of this reaction is given by,

\begin{equation}
\Gamma = \frac{g^2 m_{\sigma}}{8 \pi} \label{decaywidth}
\end{equation}

The reheating temperature is thus given by,

\begin{equation}
T_{rh} \approx 0.78 \hspace{1mm} g_{*}^{- \frac{1}{4}} \sqrt{M_{pl} \Gamma} \label{Trh1}
\end{equation}

Using equations (\ref{decaywidth}) and (\ref{Trh1}) ,

\begin{equation}
T_{rh} 
\approx 0.78 \hspace{1mm} g_{*}^{- \frac{1}{4}} \sqrt{\frac{m_{\sigma} M_{pl}}{8 \pi}} \times g \label{Trh2}
\end{equation}

The curvaton field decays at a time when $\Gamma_ {\sigma} = H$ and then

\begin{equation}
\frac{\Gamma_{\sigma}}{H_{kin}} = \frac{a_{kin}^3}{a_d^3}
\Rightarrow \frac{a_{kin}}{a_d} = \sqrt[3]{\frac{\Gamma_{\sigma}}{H_{kin}}} \label{akinuponad}         
\end{equation}

\subsection{Obtaining the Constraint}

We now derive the bound on the reheating temperature that is consistent with the nucleosynthesis constraint using this mechanism.
This reduces to finding those values of the coupling constant $g$ for which the desired reheating temperature
is obtained. These values of $g$ are obtained by deriving a bound on $g$ and it gives us the permissible values of the
coupling constant $g$ for our model.
While deriving the bound in equation (\ref{constraintonsigmasquare}) only the properties of the
Curvaton field was used. The properties of the inflaton field have not been considered so far. 
Thus apart from the Curvaton model this result is model independent. We can thus apply it to our model of
Quintessential Inflation. The above bound combined with the
fact that the curvaton energy should also be sub-dominant at the end of inflation compared with the energy of the 
inflaton field gives us a bound on the curvaton mass $m_{\sigma}$. This then combined with the required reheating
temperature gives us a bound on g where we use equation (\ref{Trh2}). 

We have from equation (\ref{Uf}),

\begin{equation}
U_f = \frac{1}{2} m_{\sigma}^2 \sigma_i^2 \label{Uf2}
\end{equation}

and from equation (\ref{dimensionlesspotential})

\begin{equation}
V_f = \frac{V_0}{\cosh \left( \beta^n \chi_f^n \right)} \label{Vf}  
\end{equation}

Dividing (\ref{Uf2}) by (\ref{Vf}),

\begin{equation}
\frac{U_f}{V_f} = \frac{1}{2} \frac{m_{\sigma}^2}{V_0}  
\cosh \left( \beta^n \chi_f^n \right) \sigma_i^2   \label{ratioofUfbyVf}
\end{equation}

Now using the bound on $\sigma_i$ given by equation (\ref{constraintonsigmasquare}) in equation 
(\ref{ratioofUfbyVf}) one can obtain,

\begin{equation}
\frac{U_f}{V_f} \ll \frac{3}{8 \pi}  \frac{m_{\sigma}^2 m_{pl}^2}{V_0} 
\cosh \left( \beta^n \chi_f^n \right) \label{ratioofUfbyVf2} 
\end{equation}

Now $\frac{U_f}{V_f} \ll 1$ implies the strong condition that,

\begin{equation}
\frac{3}{8 \pi}  \frac{m_{\sigma}^2 m_{pl}^2}{V_0} \cosh \left( \beta^n \chi_f^n \right) \ll 1  
\end{equation}

Then,

\begin{equation}
m_{\sigma}^{\frac{1}{2}} \ll 
\Big[ \frac{8 \pi}{3} \frac{1}{m_{pl}^2} \frac{V_0}{\cosh \left( \beta^n \chi_f^n \right)} \Big]^{\frac{1}{4}} \label{ccm} 
\end{equation}

We have now obtained the bound on the mass of the curvaton field. Using equation (\ref{ccm}) 
in equation (\ref{Trh2}),

\begin{equation}
\\T_{rh} \ll 0.78 g_{*}^{- \frac{1}{4}} g \frac{1}{3^{\frac{1}{4}}} \frac{1}{(8 \pi)^{\frac{1}{2}}} 
\Big[ \frac{V_0}{\cosh \left( \beta^n \chi_f^n \right)} \Big]^{\frac{1}{4}} \label{Trhbound1}
\end{equation}

We have,

\begin{equation}
T_{rh} \geq 2.2 \times 10^{12} \mbox{GeV} \label{Trhbound2}
\end{equation}

Combining equations (\ref{Trhbound1}) and (\ref{Trhbound2}),

\begin{widetext}
\begin{equation}
2.2 \times 10^{12} \mbox{GeV} \leq T_{rh}
\ll 0.78 g_{*}^{- \frac{1}{4}} g \frac{1}{3^{\frac{1}{4}}} \frac{1}{(8 \pi)^{\frac{1}{2}}} 
\Big[ \frac{V_0}{\cosh \left( \beta^n \chi_f^n \right)} \Big]^{\frac{1}{4}}
\end{equation}
\end{widetext}

Now using the constraints on the inflaton field,
$n = 6$, $\beta = 1$, $V_0 = 4.64 \times 10^{60} \mbox{GeV}$, $\chi_f = 0.88$
together with the value of $g_*$, $g_* =140$, one obtains

\begin{equation}
2.2 \times 10^{12} \mbox{GeV} \leq T_{rh} \ll 4.91577 \times 10^{13} g \hspace{1mm} \mbox{GeV} 
\end{equation}

We thus obtain the following bound on $g$,

\begin{equation}
g \gg 0.0447539 
\end{equation}

\section{Exploring the Parameter Space}

We study here two cases -

\begin{enumerate}
 \item curvaton decays before domination
 \item curvaton decays after domination
\end{enumerate}

\subsection{Curvaton Decay before Domination}                                                                

 \begin{figure}[h]
 \centering
 \includegraphics[scale=.6]{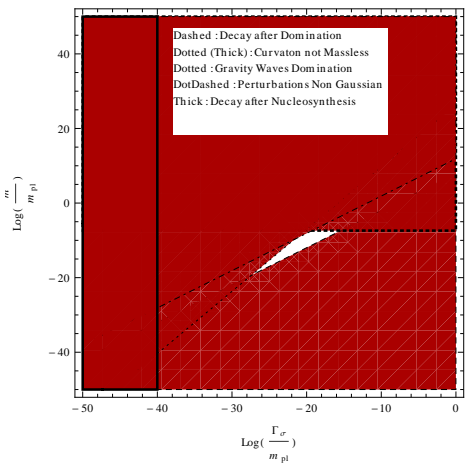}
 \caption{Curvaton constraints for the case in which the 
 curvaton field decays before domination, using N = 60 in 
 Eqs. (\ref{constraint1}), (\ref{constraint2}), (\ref{constraint3}), (\ref{constraint4}), (\ref{constraint5}),
 (\ref{constraint6}), (\ref{constraint7})
 The regions excluded by each constraint is shaded, and the allowed region is 
 unshaded.}
\label{PSDCP}
\end{figure}

We now study the case of curvaton decay before domination as applied to the same inverse hyperbolic cosine potential.
There comes a time when the energy density of the  stiff scalar matter equals the energy
density of the curvaton matter. During this moment, 

\begin{equation}
\frac{\rho_{\sigma}}{\rho_{\phi}} \Big|_{a = a_{eq}} 
= \frac{4 \pi}{3} \frac{m^2 \sigma_i^2}{m_{pl}^2 H_{kin}^2} \frac{a_m^3}{a_{kin}^3} \frac{a_{eq}^3}{a_{kin}^3} 
= 1 \label{ratioofrhos2}
\end{equation}

This is an equation that is model independent in the sense that it only depends on the 
properties of the curvaton field and not of the inflaton field. This means that we can use this equation for our potential 
directly. Using the above equation one obtains,

\begin{equation}
H_{eq} = H_{kin} \hspace{1mm} \frac{a_{kin}^3}{a_{eq}^3} 
= \frac{4 \pi}{3} \hspace{1mm} \frac{\sigma_i^2}{m_{pl}^2} \hspace{1mm} m \label{Heq}
\end{equation}

Now in order that the curvaton field should not dominate the expansion history of the Universe and that it should
decay after it becomes massive the constraint  
$H_{eq} < \Gamma_{\sigma} < m$ needs to be satisfied and thus from equation (\ref{Heq}),

\begin{equation}
\frac{4 \pi}{3} \frac{\sigma_i^2}{m_{pl}^2} m < \Gamma_{\sigma} < m \label{decaybeforedomination}
\end{equation}

Just like equation (\ref{ratioofrhos2}), equation (\ref{decaybeforedomination}) 
can also be applied to our potential directly. 

The Bardeen parameter in this case is given by,

\begin{equation}
P_{\zeta} = \frac{r_d^2}{36 \pi^2} \frac{H_i^2}{\sigma_i^2} \label{Bardeenparameter}
\end{equation}

The normalization factor indicates that the dominant component at that time is the scalar stiff matter.
Unlike in equation (\ref{ratioofrhos2}) the energy density of the curvaton field does not equal that of stiff scalar matter
and $r_d$ represents the ratio of these two densities at the exact moment when the curvaton decays.    
Here $r_d$ is given by,

\begin{equation}
r_d = \frac{\rho_{\sigma}}{\rho_{\phi}} \Big|_{a = a_d} 
= \frac{4 \pi}{3} \frac{m}{\Gamma_{\sigma}} \frac{\sigma_i^2}{m_{pl}^2} \label{rd}   
\end{equation}

Combining equations (\ref{Bardeenparameter}) and (\ref{rd}),

\begin{equation}
\frac{\sigma_i^2}{m_{pl}^2} 
= \frac{81}{4} P_{\zeta} \frac{m_{pl}^2}{H_i^2} \frac{\Gamma_{\sigma}^2}{m^2} \label{sigmauponmplwholesq}
\end{equation}

Now substituting equation (\ref{sigmauponmplwholesq}) in (\ref{decaybeforedomination}),

\begin{equation}
\Big( 27 \pi P_{\zeta} \frac{m_{pl}^2}{H_i^2} \frac{\Gamma_{\sigma}^2}{m^2} \Big) m
< \Gamma_{\sigma} < m \label{decaybeforedomination2}
\end{equation}

Define,

\begin{equation}
A = 27 \pi  P_{\zeta} \frac{m_{pl}^2}{H_i^2} = 27 \pi  P_{\zeta} \frac{m_{pl}^2}{H_{in}^2} 
\end{equation}

Fitting our model now,
$H_{in}^2 = 2.68512 \times 10^{23} GeV^2$, 
$m_{pl} = 1.220910 \times 10^{19} GeV$ and as observed from COBE 
$P_{\zeta} = 2.3 \times 10^{-9}$, one obtains

\begin{equation}
A = 1.08304 \times 10^8 \label{valueofA}
\end{equation}

From equation (\ref{decaybeforedomination2}), we have three constraints for this model.

\begin{eqnarray}
&\frac{m}{m_{pl}} > A \frac{\Gamma_{\sigma}}{m_{pl}} \\
&\frac{m}{m_{pl}} > \sqrt{A} \frac{\Gamma_{\sigma}}{m_{pl}} \\
&\frac{m}{m_{pl}} > \frac{\Gamma_{\sigma}}{m_{pl}} 
\end{eqnarray}

Thus the first three constraints for our model are,

\begin{eqnarray}
&\frac{m}{m_{pl}} > \left( 1.08304 \times 10^{8} \right) \frac{\Gamma_{\sigma}}{m_{pl}}  \label{constraint1} \\
&\frac{m}{m_{pl}} > \left( 1.04069 \times 10^{4} \right) \frac{\Gamma_{\sigma}}{m_{pl}}  \label{constraint2} \\ 
&\frac{m}{m_{pl}} > \frac{\Gamma_{\sigma}}{m_{pl}} \label{constraint3} 
\end{eqnarray}

where we have used equation (\ref{valueofA}). 

Now from equation (\ref{ratioofUfbyVf2}),

\begin{equation}
\frac{U_f}{V_f} \ll \frac{3}{8 \pi} \frac{m^2 m_{pl}^2}{V_0} \cosh \left( \beta^n \chi_f^n \right) \ll 1 
\end{equation}

which gives us,

\begin{equation}
\frac{m}{m_{pl}} \ll 
\frac{1}{m_{pl}^2}
\Big[ \frac{8 \pi}{3} \Big( \frac{V_0}{\cosh \left( \beta^n \chi_f^n \right)} \Big) \Big]^{\frac{1}{2}}
\end{equation}

Fitting our model,

\begin{equation}
\frac{m}{m_{pl}} \ll 3.97037 \times 10^{-8} \label{constraint4}
\end{equation}

This is the fourth constraint for our model.

Radiation equals the stiff scalar matter $\left( \rho_r^{(\sigma)} = \rho_{\phi} \right)$ 
at a time given by,

\begin{equation}
\frac{4 \pi}{3} \frac{m^2 \sigma_i^2}{m_{pl}^2 H_{kin}^2} \frac{a_m^3}{a_{kin}^3}
\frac{a_{eq}^2}{a_{kin}^2} \frac{a_d}{a_{kin}} = 1 \label{unity}
\end{equation}

Then using equations  (\ref{ratioofrhos}), (\ref{whencurvatonbecomesmassive}),  (\ref{akinuponad}), (\ref{unity}), 

\begin{equation}
\frac{\rho_g}{\rho_r} \Big|_{a = a_{eq}}
= \frac{16}{\pi^2} h_{GW}^2 \sqrt[3]{\frac{\Gamma_{\sigma}}{H_{kin}}}
\frac{m}{H_{kin}} \frac{m_{pl}^2}{\sigma_i^2}
\times \frac{H_{kin}^2}{m^2} 
\end{equation}

We have,

\begin{equation}
\frac{\rho_g}{\rho_r} \Big|_{a = a_{eq}}
= \frac{16}{\pi^2} h_{GW}^2 \sqrt[3]{\frac{\Gamma_{\sigma}}{H_{kin}}}
\frac{m_{pl}^2}{\sigma_i^2}
\times \frac{H_{kin}}{m} 
\ll 1
\end{equation}

Now upon using equation (\ref{sigmauponmplwholesq}),

\begin{equation}
\frac{64}{81 \pi^2 P_{\zeta}} h_{GW}^2 \frac{\Gamma_{\sigma}^{\frac{1}{3}}}{H_{kin}^{\frac{1}{3}}} 
 \frac{H_i^2}{m_{pl}^2} \frac{m^2}{\Gamma_{\sigma}^2}  \times \frac{H_{kin}}{m} \ll 1
\end{equation}

Simplifying and using $H_{kin} \approx H_{end}$,

\begin{equation}
\frac{m}{m_{pl}} \ll \frac{81 \pi^2 P_{\zeta}}{64 h_{GW}^2} 
\Big( \frac{H_{end}}{H_i^2} \Big)^2
\times \Big( \frac{m_{pl}}{H_{end}} \Big)^{\frac{8}{3}} 
\times \Big( \frac{\Gamma_{\sigma}}{m_{pl}} \Big)^{\frac{5}{3}} 
\end{equation}

Also using , $P_{\zeta} = 2.3 \times 10^{-9}$, $h_{GW}^2 = 5.82139 \times 10^{-15}$, 
$m_{pl} = 1.220910 \times 10^{19} GeV$,
$H_{in} = 5.18181 \times 10^{11} \mbox{GeV}$ and $H_{end} = 6.02438 \times 10^{11} GeV$  
we obtain the 5th constraint for the model,

\begin{equation}
\frac{m}{m_{pl}} 
\ll 
2.03652 \times 10^{26} \Big( \frac{\Gamma_{\sigma}}{m_{pl}} \Big)^{\frac{5}{3}} \label{constraint5}
\end{equation}

The curvaton perturbations should satisfy the Gaussianity condition.
This condition does not hold automatically.
By imposing this condition  we restrict the amplitude of the perturbations which are then
negligible compared to the mean value of the curvaton field,

\begin{equation}
\sigma_i^2 \gg \frac{H_i^2}{4 \pi^2} 
\end{equation}

We now divide by $m_{pl}^2$, use equation (\ref{sigmauponmplwholesq}) and simplify to obtain,  

\begin{equation}
\frac{m}{m_{pl}} \ll 9 \pi \sqrt{P_{\zeta}} \times \frac{m_{pl}^2}{H_i^2}
\Big( \frac{\Gamma_{\sigma}}{m_{pl}} \Big)
\end{equation}

Now $P_{\zeta} = 2.3 \times 10^{-9}$, $m_{pl} = 1.220910 \times 10^{19} GeV$ and
$H_{in} = 5.18181 \times 10^{11} GeV$

\begin{equation}
\frac{m}{m_{pl}} 
\ll 
\left( 7.52766 \times 10^{11} \right) \hspace{1mm} \frac{\Gamma_{\sigma}}{m_{pl}} \label{constraint6}
\end{equation}

We thus obtain the sixth constraint for our model. 
At nucleosynthesis things should now proceed according to the standard Big Bang scenario
and hence we also have a seventh constraint for our model,     

\begin{equation}
\Gamma_{\sigma} > 10^{-40} m_{pl} 
\end{equation}

or

\begin{equation}
\frac{\Gamma_{\sigma}}{m_{pl}} > 10^{-40} \label{constraint7}
\end{equation}

We plot the allowed region in parameter space as permitted by these constraints in Fig.1. While plotting this 
figure we take the number of e foldings, $N = 60$, $n = 6$ $\beta = 1$, $V_0 = 4.64 \times 10^{60} \mbox{GeV}$ and
$\chi_{f} = 0.88$, which are constraints on the inflaton field. Out of the constraints given by equations. 
(\ref{constraint1} $-$ \ref{constraint3})
only the strongest one is shown in the figure whereas the remaining are automatically satisfied. We
then plot the constraints given by equations.
(\ref{constraint4}), (\ref{constraint5}), (\ref{constraint6}) and (\ref{constraint7}).

\subsection{Curvaton Decay after Domination}

\begin{figure}[h]
 \centering
 \includegraphics[scale=.6]{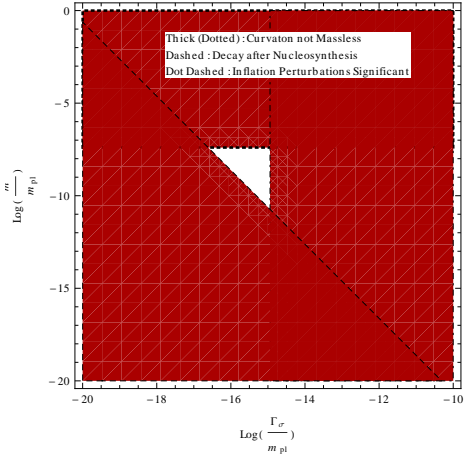}
 \caption{Curvaton constraints for the case in which the 
 curvaton field decays after domination, using N = 60 in 
 Eqs. (\ref{constraint1dad}), (\ref{constraint2dad}), (\ref{constraint3dad}), 
 (\ref{constraint4dad}), (\ref{constraint5dad}) 
 The regions excluded by each constraint is shaded, and the allowed region is 
 unshaded.}
\label{PSDCP}
\end{figure}

Equation (\ref{constraintonsigmasquare}) gives our first constraint,

\begin{equation}
\frac{\sigma_i}{m_{pl}} \ll 0.488603 \label{constraint1dad}
\end{equation}

Equation (\ref{constraint4}),

\begin{equation}
\frac{m}{m_{pl}} \ll 3.97037 \times 10^{-8} \label{constraint2dad}
\end{equation}

gives us our second constraint.

We desire that the curvaton field should decay before nucleosynthesis occurs. Then the following inequality is satisfied,
$H_{\mbox{nucl}} = 10^{-40} m_{pl} < \Gamma_{\sigma}$. 
The curvaton should also decay after it dominates the expansion of the Universe, and
then we have the following inequality 
$\Gamma_{\sigma} < H_{eq} = \frac{4 \pi}{3} \frac{\sigma_i^2}{m_{pl}^2} m$. Thus we obtain,

\begin{equation}
10^{-40} m_{pl} < \Gamma_{\sigma} < \frac{4 \pi}{3} \frac{\sigma_i^2}{m_{pl}^2} m 
\end{equation}

This gives us our third constraint,

\begin{equation}
\frac{m}{m_{pl}} \hspace{1mm} \frac{\sigma_i^2}{m_{pl}^2} \gg \frac{3}{4 \pi} \times 10^{-40} \label{constraint3dad}  
\end{equation}

We now require that gravity waves do not dominate,

\begin{equation}
\frac{\rho_g}{\rho_{\sigma}} \Big|_{a = a_{eq}} 
= \frac{64}{3 \pi} h_{GW}^2 
\Big( \frac{3}{4 \pi} \frac{m_{pl}^2}{\sigma_i^2}  \frac{H_{kin}}{m} \Big)^{\frac{2}{3}} \ll 1 \label{ratioofrhos3}
\end{equation}

When decay occurs after curvaton domination, the produced perturbation is, 

\begin{equation}
P_{\zeta} = \frac{1}{9 \pi^2} \frac{H_{in}^2}{\sigma_i^2} \label{Bardeenparameter2}
\end{equation}

Here $P_{\zeta}$ is the spectrum of the Bardeen parameter.

Now using equations  
(\ref{ratioofrhos3}), (\ref{Bardeenparameter2}), $h_{GW}^2 = \frac{H_{in}^2}{8 \pi M_{pl}^2}$ 
and $H_{kin} \approx H_{end}$, we obtain,

\begin{equation}
\frac{m}{m_{pl}} \gg 
\Big[ \frac{(192 \pi P_{\zeta})^{\frac{3}{2}} \times 3 \times H_{end}}{4 \pi m_{pl}} \Big]
\frac{\sigma_i}{m_{pl}}
\end{equation}

giving us our fourth constraint,

\begin{equation}
\frac{m}{m_{pl}} \gg \Big( 1.9249 \times 10^{-17} \Big) \frac{\sigma_i}{m_{pl}} \label{constraint4dad}
\end{equation}

The fluctuations of the inflaton field are measured by a quantity called the power spectrum,

\begin{equation}
P_{\phi} = \Big( \frac{H}{2 \pi} \Big)^2 
\end{equation}

Now the inflaton field fluctuations should be insignificant compared to those of the curvaton field. Then the 
power spectrum $P_{\phi}$ should be less than unity,

\begin{equation}
P_{\phi} \ll 1  
\end{equation}

Thus,

\begin{equation}
\Big( P_{\phi} \Big)_i \ll 1 
\end{equation}

This gives us,

\begin{equation}
\frac{\sigma_i}{m_{pl}} \ll \frac{2}{3 m_{pl} \sqrt{P_{\zeta}}} 
\end{equation}

upon using equation (\ref{Bardeenparameter2}).

Now using, $m_{pl} = 1.220910 \times 10^{19} GeV/c^2$ and $P_{\zeta} = 2.3 \times 10^{-9}$ we obtain our 5th
constraint,

\begin{equation}
\Big( \frac{\sigma_i}{m_{pl}} \Big) \ll 1.13857 \times 10^{-15} \label{constraint5dad}
\end{equation}

We plot the allowed region in parameter space as permitted by these constraints in figure 2.
While plotting this 
figure we take the number of e foldings, $N = 60$, $n = 6$ $\beta = 1$, $V_0 = 4.64 \times 10^{60} \mbox{GeV}$ and
$\chi_{f} = 0.88$, which are constraints on the inflaton field.
We then show the constraints given by
eqns. (\ref{constraint1dad}), (\ref{constraint2dad}), (\ref{constraint3dad}), (\ref{constraint4dad})
and (\ref{constraint5dad}). Out of these, two are redundant and we plot the other three. 

\section{Conclusion}

We study a single scalar field model 
of quintessential inflation with inverse cosh hyperbolic potential 
where the reheating temperature is obtained using the curvaton mechanism. The temperature thus obtained is consistent with the
nucleosynthesis constraint,$T_{rh} \gtrsim 2.2 \times 10^{12} \mbox{GeV}$ due to relic gravity waves.
We find that upon implementing this mechanism,
$2.2 \times 10^{12} \mbox{GeV} \leq T_{rh} \ll 4.91577 \times 10^{13} g \hspace{1mm} \mbox{GeV} $. The upper bound on temperature translates into a bound on the coupling constant g, namely,
$g \gg 0.0447539$. The nucleosynthesis constraint is thus satisfied. 

We have studied the cases of curvaton decay before domination
and curvaton decay after domination
and plotted the allowed region in parameter space. The 
case for decay before domination is given in Figure 1 and that for decay after domination is given in Figure 2.
While plotting these 
figures we take the number of e-foldings, $N = 60$, $n = 6$ $\beta = 1$, $V_0 = 4.64 \times 10^{60} \mbox{GeV}$ and
$\chi_{f} = 0.88$, which represent constraints on the inflaton field.
Out of the constraints given by equations. 
(\ref{constraint1} $-$ \ref{constraint3}) only the strongest one is shown in the Fig.1       
whereas the remaining two are automatically satisfied. We
then plot the constraints given by Eqs.
(\ref{constraint4}), (\ref{constraint5}), (\ref{constraint6}) and (\ref{constraint7}).
While plotting Fig.2 we show the constraints given by
eqns. (\ref{constraint1dad}), (\ref{constraint2dad}), (\ref{constraint3dad}), (\ref{constraint4dad})
and (\ref{constraint5dad}). Out of these, two are redundant and we plot the other three. 
While plotting these figures the regions excluded by each constraint is shaded, and the allowed region is
unshaded. From Fig. 1 and Fig. 2 it is clear that the allowed region in parameter space is very small and that real situations
arising in both the cases are severely restricted by many constraints.

\bibliography{CB}

\begin{thebibliography}{10}

\bibitem{Abhineet}
A.~{Agarwal}, R.~{Myrzakulov}, M.~{Sami}, and N.~K. {Singh},
\newblock Physics Letters B {\bf 770}, 200 (2017), 1708.00156.

\bibitem{CCIsBack}
L.~M. {Krauss} and M.~S. {Turner},
\newblock General Relativity and Gravitation {\bf 27}, 1137 (1995),
  astro-ph/9504003.

\bibitem{GuthInflation}
A.~H. Guth,
\newblock Phys. Rev. D {\bf 23}, 347 (1981).

\bibitem{Wali1}
M.~{Wali Hossain},
\newblock ArXiv e-prints  (2018), 1801.03272.

\bibitem{Wali2}
M.~W. {Hossain}, R.~{Myrzakulov}, M.~{Sami}, and E.~N. {Saridakis},
\newblock International Journal of Modern Physics D {\bf 24}, 1530014 (2015),
  1410.6100.

\bibitem{WoodOwen}
K.~{Dimopoulos}, L.~D. {Wood}, and C.~{Owen},
\newblock \prd {\bf 97}, 063525 (2018), 1712.01760.

\bibitem{QIPeebles}
P.~J.~E. {Peebles} and A.~{Vilenkin},
\newblock \prd {\bf 59}, 063505 (1999), astro-ph/9810509.

\bibitem{SineHyperbolicInflation}
S.~{Basilakos} and J.~D. {Barrow},
\newblock \prd {\bf 91}, 103517 (2015), 1504.03469.

\bibitem{CurvatonBrane}
A.~R. {Liddle} and L.~A. {Ure{\~n}a-L{\'o}pez},
\newblock \prd {\bf 68}, 043517 (2003), astro-ph/0302054.

\bibitem{LythUngarelliWands}
D.~H. {Lyth}, C.~{Ungarelli}, and D.~{Wands},
\newblock \prd {\bf 67}, 023503 (2003), astro-ph/0208055.

\bibitem{LythWands}
D.~H. {Lyth} and D.~{Wands},
\newblock Physics Letters B {\bf 524}, 5 (2002), hep-ph/0110002.

\bibitem{FengLi}
B.~{Feng} and M.~{Li},
\newblock Physics Letters B {\bf 564}, 169 (2003), hep-ph/0212213.

\bibitem{BartoloLiddle}
N.~{Bartolo} and A.~R. {Liddle},
\newblock \prd {\bf 65}, 121301 (2002), astro-ph/0203076.

\bibitem{MartinSloth}
M.~S. Sloth,
\newblock Nucl. Phys. {\bf B656}, 239 (2003), hep-ph/0208241.

\bibitem{ByrnesCortLiddle}
C.~T. {Byrnes}, M.~{Cort{\^e}s}, and A.~R. {Liddle},
\newblock \prd {\bf 90}, 023523 (2014), 1403.4591.

\bibitem{MazumdarRocher}
A.~{Mazumdar} and J.~{Rocher},
\newblock PHYSREP {\bf 497}, 85 (2011), 1001.0993.

\bibitem{EnqvistLernerTaanila}
K.~{Enqvist}, R.~N. {Lerner}, and O.~{Taanila},
\newblock JCAP {\bf 12}, 016 (2011), 1105.0498.

\bibitem{GongKitajimaTerada}
J.-O. {Gong}, N.~{Kitajima}, and T.~{Terada},
\newblock JCAP {\bf 3}, 053 (2017), 1611.08975.

\bibitem{ChingangbamHuang}
P.~{Chingangbam} and Q.-G. {Huang},
\newblock \prd {\bf 83}, 023527 (2011), 1006.4006.

\bibitem{InstantPreheating}
G.~N. Felder, L.~Kofman, and A.~D. Linde,
\newblock Phys. Rev. {\bf D59}, 123523 (1999), hep-ph/9812289.

\bibitem{InstantPreheating2}
G.~N. Felder, L.~Kofman, and A.~D. Linde,
\newblock Phys. Rev. {\bf D60}, 103505 (1999), hep-ph/9903350.

\bibitem{InflationafterPreheating}
G.~N. Felder, L.~Kofman, A.~D. Linde, and I.~Tkachev,
\newblock JHEP {\bf 08}, 010 (2000), hep-ph/0004024.

\bibitem{Campos1}
A.~H. Campos, J.~M.~F. Maia, and R.~Rosenfeld,
\newblock Phys. Rev. {\bf D70}, 023003 (2004), astro-ph/0402413.

\bibitem{Campos2}
A.~H. Campos, H.~C. Reis, and R.~Rosenfeld,
\newblock Phys. Lett. {\bf B575}, 151 (2003), hep-ph/0210152.

\bibitem{FermionPreheating}
S.~Tsujikawa, B.~A. Bassett, and F.~Viniegra,
\newblock JHEP {\bf 08}, 019 (2000), hep-ph/0006354.

\bibitem{Starobinsky}
L.~{Kofman}, A.~{Linde}, and A.~A. {Starobinsky},
\newblock \prd {\bf 56}, 3258 (1997), hep-ph/9704452.

\bibitem{Starobinsky2}
L.~{Kofman}, A.~{Linde}, and A.~A. {Starobinsky},
\newblock Physical Review Letters {\bf 73}, 3195 (1994), hep-th/9405187.

\bibitem{DimopoulosEtaProblem}
K.~{Dimopoulos},
\newblock \prd {\bf 68}, 123506 (2003), astro-ph/0212264.

\bibitem{DimopoulosLyth}
K.~{Dimopoulos} and D.~H. {Lyth},
\newblock \prd {\bf 69}, 123509 (2004), hep-ph/0209180.

\bibitem{KLTY}
N.~{Kitajima}, D.~{Langlois}, T.~{Takahashi}, and S.~{Yokoyama},
\newblock JCAP {\bf 12}, 042 (2017), 1707.06929.

\bibitem{TorradoByrnesHardwickVenninWands}
J.~{Torrado}, C.~T. {Byrnes}, R.~J. {Hardwick}, V.~{Vennin}, and D.~{Wands},
\newblock ArXiv e-prints  (2017), 1712.05364.

\bibitem{QIPlanck2015SamiWali}
C.-Q. {Geng}, M.~W. {Hossain}, R.~{Myrzakulov}, M.~{Sami}, and E.~N.
  {Saridakis},
\newblock \prd {\bf 92}, 023522 (2015), 1502.03597.

\bibitem{LythontheHilltop}
L.~{Boubekeur} and D.~H. {Lyth},
\newblock JCAP {\bf 7}, 010 (2005), hep-ph/0502047.

\bibitem{planck2015}
Planck, P.~A.~R. Ade {\em et~al.},
\newblock (2015), 1502.02114.

\end{thebibliography}

\bibliographystyle{h-physrev3}

\end{document}